\documentclass[aps,pra,twocolumn,superscriptaddress]{revtex4-1}

\usepackage{amsmath}
\usepackage{graphicx}
\usepackage{dcolumn}
\usepackage[colorlinks=true, allcolors=blue]{hyperref}
\usepackage[super]{nth}

\usepackage{siunitx}

\usepackage{textgreek}
\usepackage{float}
\begin{document}

\title{Intense beam of metastable Muonium}

\author{G.~Janka}
\author{B.~Ohayon}
\author{Z.~Burkley}
\author{L.~Gerchow}
\affiliation{
Institute for Particle Physics and Astrophysics, ETH Z\"urich, CH-8093 Z\"urich, Switzerland
}
\author{N.~Kuroda}
\affiliation{Institute of Physics, The University of Tokyo, 153-8902 Tokyo, Japan}
\author{X.~Ni}
\affiliation{Laboratory for Muon Spin Spectroscopy, Paul Scherrer Institute, CH-5232 Villigen PSI, Switzerland}
\author{R.~Nishi}
\affiliation{Institute of Physics, The University of Tokyo, 153-8902 Tokyo, Japan}
\author{Z.~Salman}
\affiliation{Laboratory for Muon Spin Spectroscopy, Paul Scherrer Institute, CH-5232 Villigen PSI, Switzerland}
\author{A.~Suter}
\affiliation{Laboratory for Muon Spin Spectroscopy, Paul Scherrer Institute, CH-5232 Villigen PSI, Switzerland}
\author{M.~Tuzi}
\author{C.~Vigo}
\affiliation{
Institute for Particle Physics and Astrophysics, ETH Z\"urich, CH-8093 Z\"urich, Switzerland
}
\author{T.~Prokscha}
\affiliation{Laboratory for Muon Spin Spectroscopy, Paul Scherrer Institute, CH-5232 Villigen PSI, Switzerland}
\author{P.~Crivelli}
 \email{crivelli@phys.ethz.ch}

\affiliation{
Institute for Particle Physics and Astrophysics, ETH Z\"urich, CH-8093 Z\"urich, Switzerland
}

\date{\today}

\begin{abstract}

Precision spectroscopy of the Muonium Lamb shift and fine structure requires a robust source of 2S Muonium. To date, the beam-foil technique is the only demonstrated method for creating such a beam in vacuum. Previous experiments using this technique were statistics limited, and new measurements would benefit tremendously from the efficient 2S production at a low energy muon ($<20$ keV) facility. Such a source of abundant low energy $\mathrm{\mu^+}$ has only become available in recent years, e.g.~at the Low-Energy Muon beamline at the Paul Scherrer Institute. Using this source, we report on the successful creation of an intense, directed beam of metastable Muonium. 
We find that even though the theoretical Muonium fraction is maximal in the low energy range of $2-5$ keV, scattering by the foil and transport characteristics of the beamline favor slightly higher $\mathrm{\mu^+}$ energies of $7-10$ keV. 
We estimate that an event detection rate of a few events per second for a future Lamb shift measurement is feasible, enabling an increase in precision by two orders of magnitude over previous determinations.


\end{abstract}

 
\maketitle


Muonium (M) is the bound state of a positive muon ($\mathrm{\mu^+}$) and an electron, two particles devoid of internal structure. Therefore, and in contrast to hydrogen, theory and experiment with M can be compared free of finite-size and nuclear effects \cite{2005-PSAS}. Testing bound state quantum electrodynamics (QED) in the muonic sector is highly motivated by the inconsistencies which have arisen there, e.g.~the deviation of the measured anomalous magnetic moment of the muon from its theoretical value \cite{2103-g2}, and the difference between the proton radius as measured by laser spectroscopy of muonic hydrogen \cite{2013-Puzzle} and several experiments in electronic hydrogen \cite{2016-CODATA14,2018-1S3S, Scattering2015}. However, the puzzle is arguably nearing its solution \cite{2018-Bayer,2019-Eric,2019-PRAD}.

To this day, precision experiments with M only utilized its ground-state \cite{2000-Jung,2016-Jung}. The 1S-2S transition was measured by pulsed laser spectroscopy \cite{1988-1S,2000-Meyer}, putting tight bounds on the muon-electron charge ratio. A future precision measurement using a CW laser is planned by the Mu-MASS experiment \cite{2018-MuMASS}. The measurement of the ground-state hyperfine structure currently determines the muon magnetic moment with the highest precision \cite{1999-HPF}, with an improvement underway by the MUSEUM collaboration \cite{2017-Museum,2019-MUseum}. However, the methods used for M production in these measurements do not produce sufficient 
M(2S) in vacuum, and so cannot be used to study transitions from long-lived excited states.
These include the $n=2$ Lamb shift \cite{1984-LS,1990-LS} and fine-structure \cite{1990-FS}, which were measured previously in M. Other transitions probed in hydrogen with fast beams may be considered as well \cite{1971-n345,1972-n3,1979-s-d,1982-FastBalmer}.

Similar to hydrogen, metastable M can be formed with the beam-foil technique \cite{1974-BFreview}, and indeed M(2S) was first observed at the TRIUMF cyclotron accelerator using sub-surface $\mathrm{\mu^+}$ at $2.1$ MeV, impinging on gold and aluminum foils \cite{1981-Mu2S}, followed by an observation at the Los Alamos Meson Physics Facility (LAMPF) \cite{1984-LAMPF-4h}. In the Born approximation, production of M with this beam-foil technique is expected to be comparable to hydrogen with protons at the same velocity \cite{1965-born}, favouring energies of several keV \cite{1981-MuVacuum,1998-Solid}. For this reason, the TRIUMF and LAMPF muon beams had to be heavily lowered in energy by degrader foils with the price of losing roughly half of the beam intensity. This resulted in a wide angular distribution for the emitted M \cite{1992-Ang}. The broad energy distribution of the degraded muon beam, extending from keVs to MeVs, resulted in a wide M energy distribution peaking at low energy and extending up to $20$ keV. This process severely limited the overall M yield.
At TRIUMF, the estimated production rate of M(2S) per incident muon was $0.08\%$ \cite{1984-LS}. This low efficiency, combined with large divergence of the beam, and a large muon-related background, resulted in a maximal detection rate of a few events per hour \cite{1985-TIUMFLSThesis}. This limited the precision of the Lamb shift measurement to $1\%$ \cite{1984-LS}. Another campaign was conducted in parallel at the LAMPF accelerator, using similar methods and arriving at a comparable statistical uncertainty of $2\%$ \cite{1990-LS}.


%
\begin{figure*}[!htbp]
\centering
\includegraphics[width=1.8\columnwidth,trim={0 10 0 10},clip]{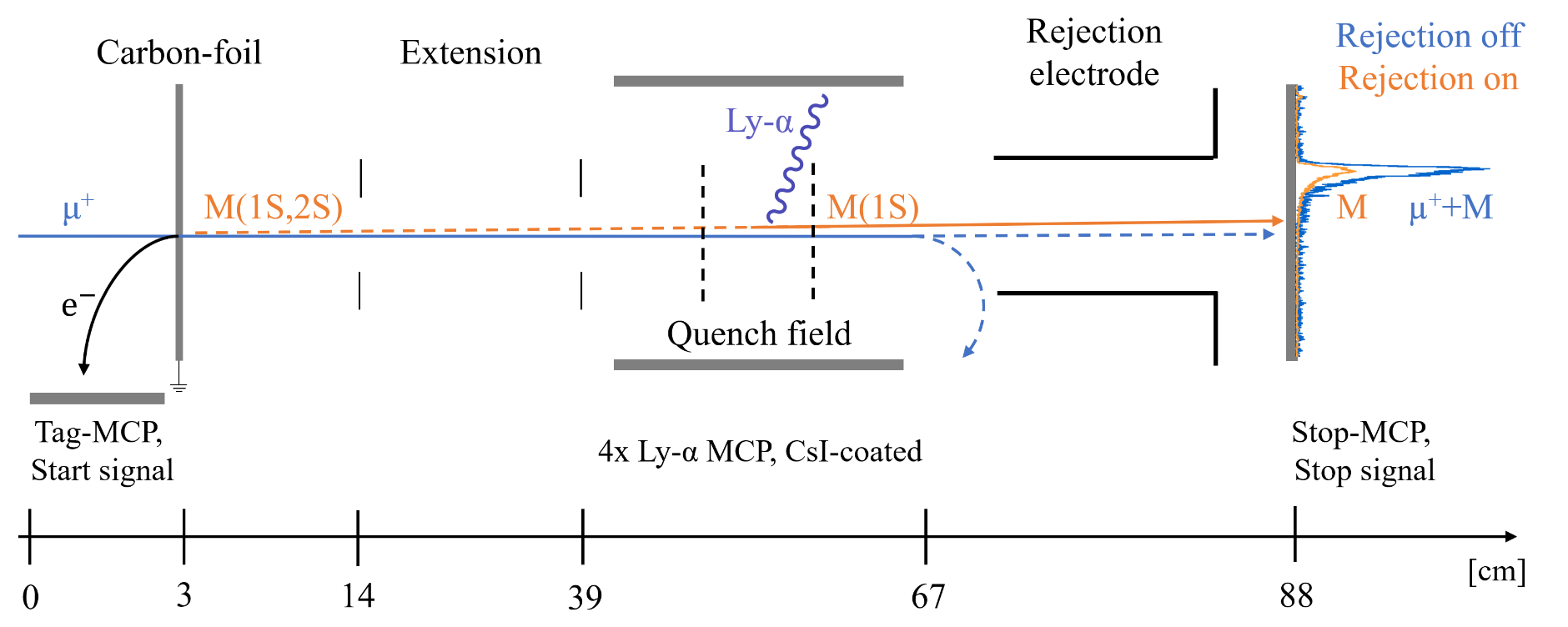}
\caption{\label{fig:beamline}The experimental setup installed at the end of the LEM beamline (see text), lengths at the bottom are not to scale. To the right is an example of the time distribution of particles reaching the Stop-MCP in coincidence with the Tag-MCP. The paths of M and $\mathrm{\mu^+}$ are not parallel for visualization reasons.}
\end{figure*}
To enable the next generation of precision measurements with metastable M, we set out to solve the main limitation affecting previous campaigns. 
%
%
In this communication, we report on the efficient creation and detection of a nearly collinear M(2S) beam by the beam-foil technique. Employing the slow $\mathrm{\mu^+}$ from the Low-Energy-Muon (LEM) beamline at the Paul Scherrer Institute (PSI) \cite{2004-LEM,2008-LEM} with energies $<$ 10 keV allowed us to directly use a thin ($\sim15$ nm) carbon foil as a conversion target without the necessity of any prior beam degradation. By tagging each muon and measuring its time-of-flight (TOF), we report for the first time the M creation efficiency over well-defined exit energies ranging from $2$ to $8$ keV. Through quenching the metastable 2S state to the short-lived 2P state in a static electric field and detecting the emitted Lyman-$\alpha$ photons, we extract the 2S fraction and compare with estimation from the literature. Combining our results with particle tracing simulations, we were able to quantify the M(2S) beam parameters. These parameters enable a realistic estimation of the achievable event rate for a future Lamb shift measurement. We conclude that a significant improvement over the state-of-the-art is within reach.




The LEM beamline at PSI generates low energy muons by moderating a surface $\mathrm{\mu^+}$ beam ($4$ MeV energy) from the $\mathrm{\mu}$E4 beamline with a silver foil coated with a thick layer of a solid noble gas mixture \cite{1987_RareGasMod,2001_Prokscha_ModGratings,2008-LEM}. For the experiment conducted here, a solid neon moderator was used, allowing the formation of a slow, monoenergetic ($2-12$ keV) $\mathrm{\mu^+}$ beam. 
In this energy regime, hydrogen formation data \cite{1981_Gabrielse,1994-Gonin} suggests a high production rate of M, some in metastable states, by impinging $\mathrm{\mu^+}$ on a foil. In the measurements performed, three different incident energies $E_\mathrm{inc}$ of $5$, $7.5$, and $10$ keV were chosen. For each $E_\mathrm{inc}$, the beamline parameters were optimized, utilizing the Geant4-based musrSim simulation \cite{2012_MusrSim, 2015-GEANT4}. Therefore, while the highest $\mathrm{\mu^+}$ to M conversion efficiencies are expected at the lowest $\mathrm{\mu^+}$ energies, we gain in transportation and detection efficiency with increasing energy. An energy of $5$ keV appeared to be a lower threshold in this regard. 

\begin{figure}[!b]
\includegraphics[width=1\columnwidth,trim={0 4 0 30},clip]{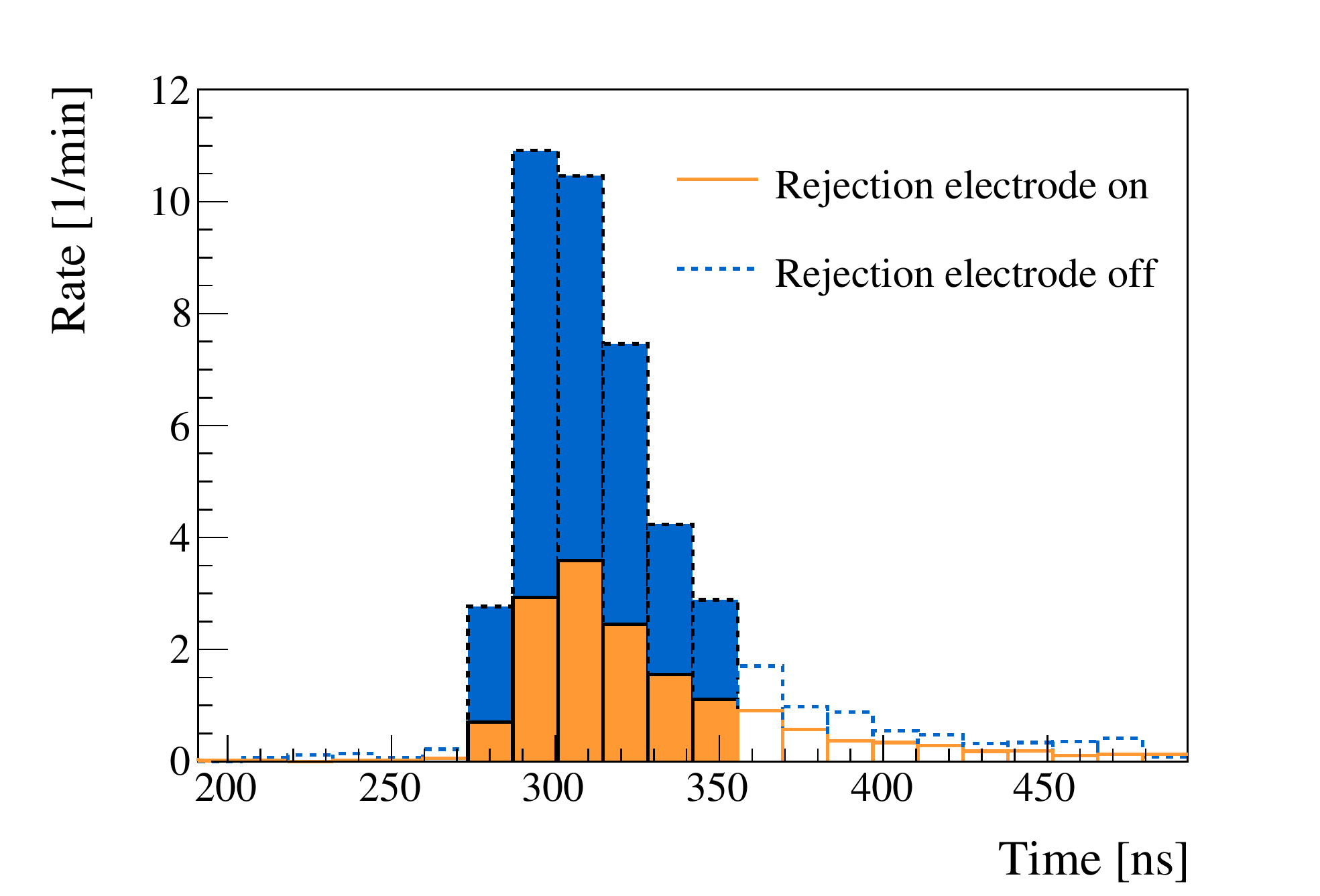}
\caption{\label{fig:mu_fraction_ext} Histograms obtained from the dataset of $10$ keV $E_\mathrm{inc}$ after background subtraction. The solid line data is with rejection field on and corresponds to pure M signal. The dotted data is with rejection electrode off and corresponds to M and $\mathrm{\mu^+}$ signal. The filled bins were used to extract M fractions, whereas the hollow bins were ignored due to large statistical uncertainty and additional background. }
\end{figure}

The experimental setup for generating and characterizing the M and M(2S) beam is shown in Fig.~\ref{fig:beamline}. Muons from the LEM beamline are directed onto a carbon foil. A fraction of the $\mu^{+}$ captures an electron while passing through the foil, forming M primarily in the ground and $n=2$ states. The carbon foil can also be used to tag incoming $\mathrm{\mu^+}$. When an incoming muon hits the foil, on average one secondary electron (SE) is released upstream \cite{2016-FoilsScat}. This SE is guided to a microchannel plate (Tag-MCP), giving the start time for the experiment.
Upon transmission, SE formed downstream might also be created. To prevent these electrons from creating false signals, each subsequent detector is biased to reject them.

The beam emerging from the foil, which is grounded to prevent quenching of any M formed in the 2S state, 
propagates in a field-free region, and then through an electrical quenching region formed by two ring electrodes that mixes the 2S and 2P states. The field at the center is $400$ V/cm. Unlike the metastable 2S state, the 2P state is short-lived ($\tau_{2P} \approx 1.6$ ns) and relaxes to the ground state within a few nanoseconds, emitting a photon of $122$ nm (Ly-$\alpha$). This photon can be detected by four CsI-coated MCPs (Ly-$\alpha$-MCP) surrounding the quenching area. The beam exiting the quenching region, now containing predominantly M(1S) and $\mathrm{\mu^+}$, reaches a rejection electrode at high voltage ($E_\mathrm{inc}$ + $1$ kV) that only allows passage of M(1S). The surviving M(1S) impinge onto an MCP (Stop-MCP), providing the stop signal.





The fraction of M formed out of the incident muon beam, $f_{\mathrm{M/\mu^+}}$, is extracted from coincidence events between the Tag- and Stop-MCP with the rejection electrode turned on or off for different $E_\mathrm{inc}$. The TOF spectra for rejection off (M+$\mathrm{\mu^+}$) and rejection on (M), after a subtraction of a constant background of $0.1$ counts/min, are divided into time bins, with the results for 10 keV incident $\mathrm{\mu^+}$ shown in Fig.~\ref{fig:mu_fraction_ext}.

\begin{figure}[!t]
\includegraphics[width=1\columnwidth,trim={0 4 0 30},clip]{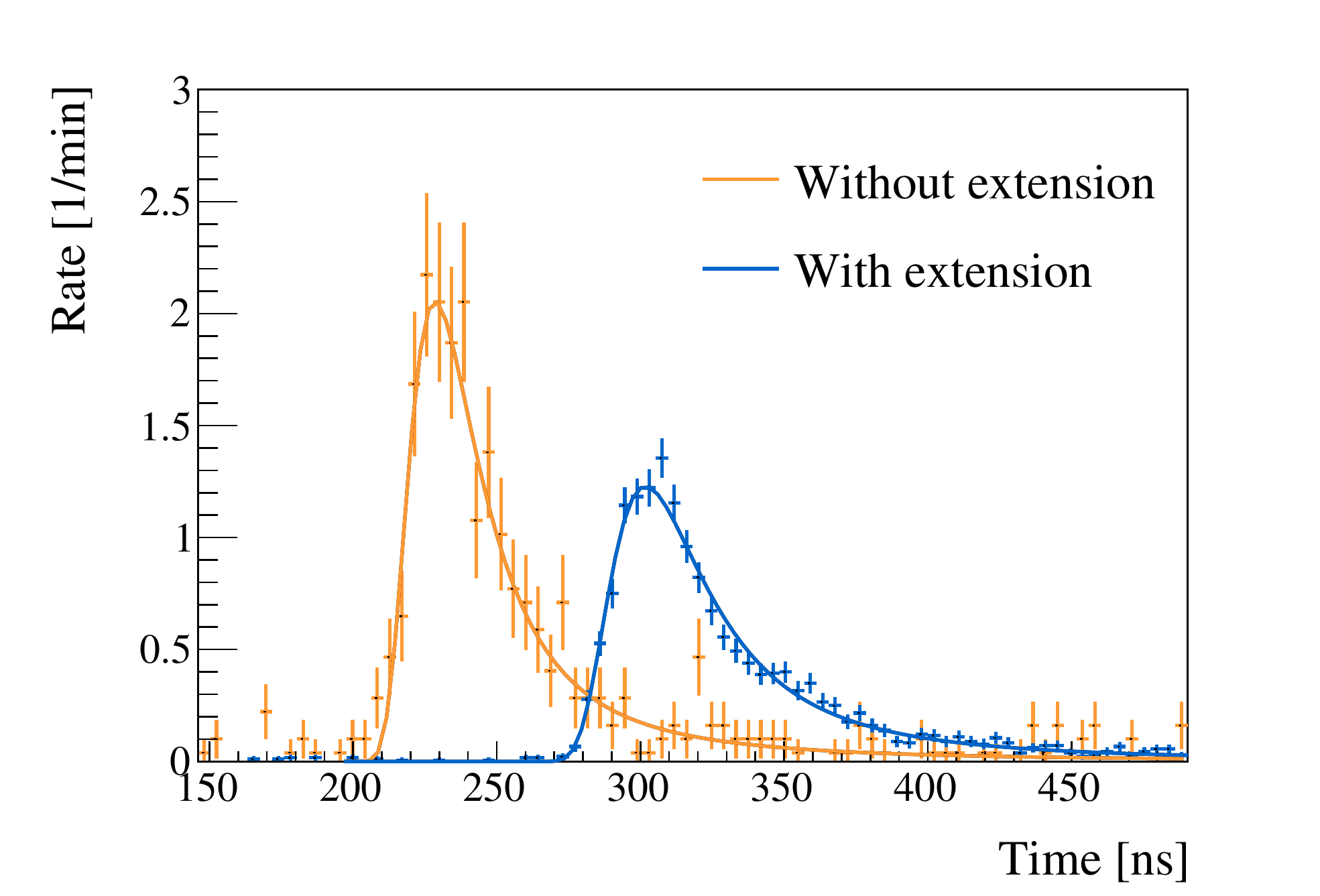}
\caption{\label{fig:t0}TOF distributions of M for $10$ keV $\mathrm{\mu^+}$ incident on the foil, with (dark blue) and without (light orange) extension stage. A Landau distribution was used for fitting the spectra.}
\end{figure}

\begin{figure}[!t]
\includegraphics[width=1\columnwidth,trim={0 4 0 30},clip]{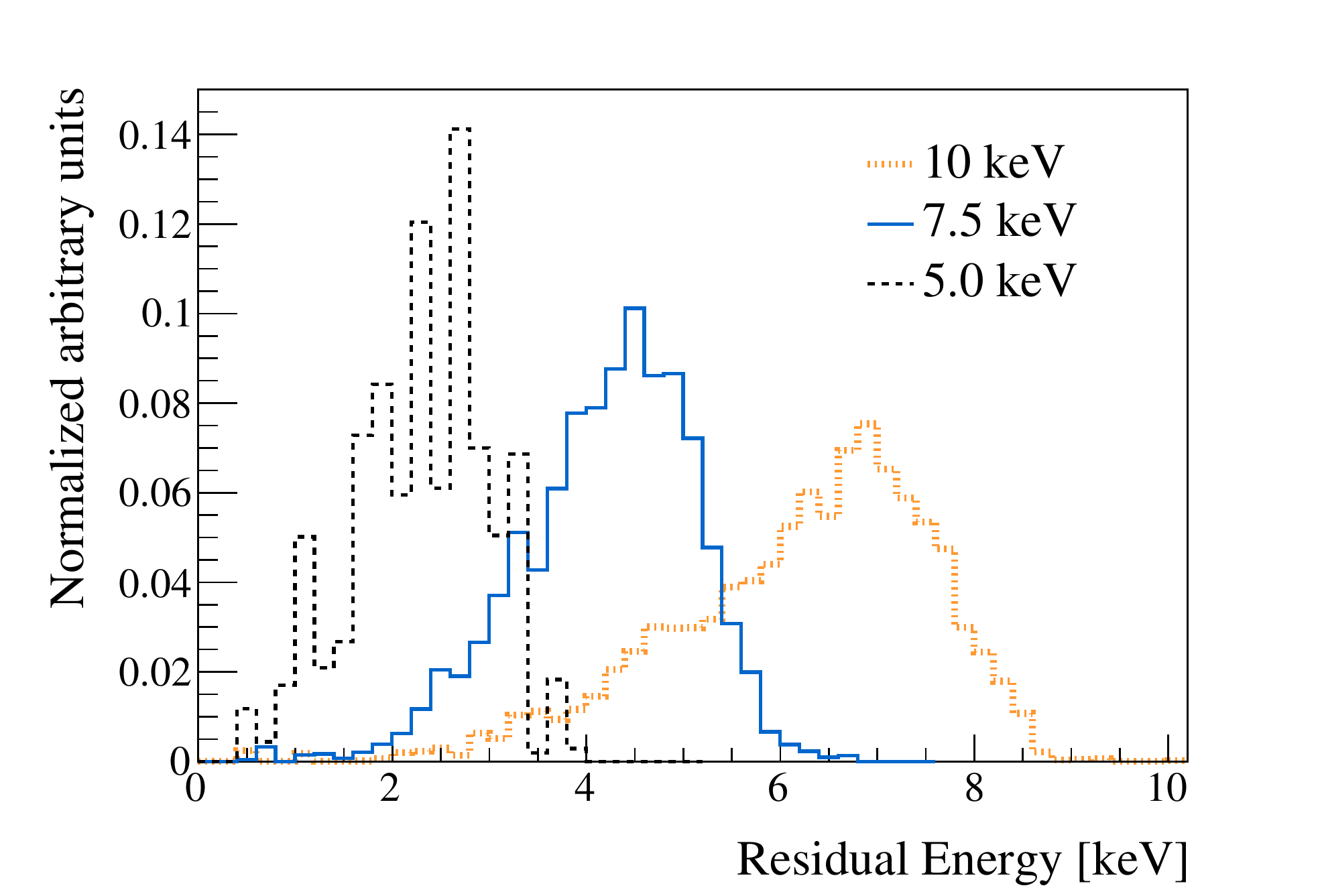}
\caption{\label{fig:energy_distribution} Energy distributions of M reaching the Stop-MCP, measured at three different $E_\mathrm{inc}$. Areas are normalized to $1$. 
}
\end{figure}

An extension stage can be added between the carbon foil and the quenching region to extend the travelling distance. The resulting increase in TOF allows the extraction of the velocity and thus energy distributions of both $\mathrm{\mu^+}$ and M after the foil, which are not known a priori.
Additionally, the extension stage ensures that all 2P states, as well as higher lying states produced in the foil \cite{1981-Hvac}, decay prior to reaching the quenching region.
For $10$ keV incident $\mathrm{\mu^+}$, the spectra were measured with and without the extension stage (Fig.~\ref{fig:t0}).
A Landau distribution was found to describe well the TOF spectra.
In addition to the length of the stage and the entire distance between foil and Stop-MCP, the time offset of the detection system was determined with a linear fit to be $t_0 = 51 \pm 4$ ns.
As the extension stage was always present during the measurements with $E_\mathrm{inc}$ of $5.0$ and $7.5$ keV, $t_0$ and the total length were used to convert these TOF spectra to the energy distributions presented in Fig.~\ref{fig:energy_distribution}.

We found that the most probable energy loss in the foil is $2.3-3.0$ keV (see Table~\ref{tab:summary}).
The foil thickness can be derived from the results for the Most Probable Energy (MPE) and the corresponding energy distributions by comparing them with the LEM Geant4 simulation, in which an effective, calibrated interaction with the foil is implemented \cite{2015-GEANT4}.
We find that a thickness of $15$ nm is most probable, which is more than the $10$ nm specified by the manufacturer.
This fact is not surprising considering the differences between the nominal and derived carbon foil thicknesses determined in \cite{2015-AllegLoss}.

\begin{figure}[!b]
\includegraphics[width=1\columnwidth,trim={0 12 0 30},clip]{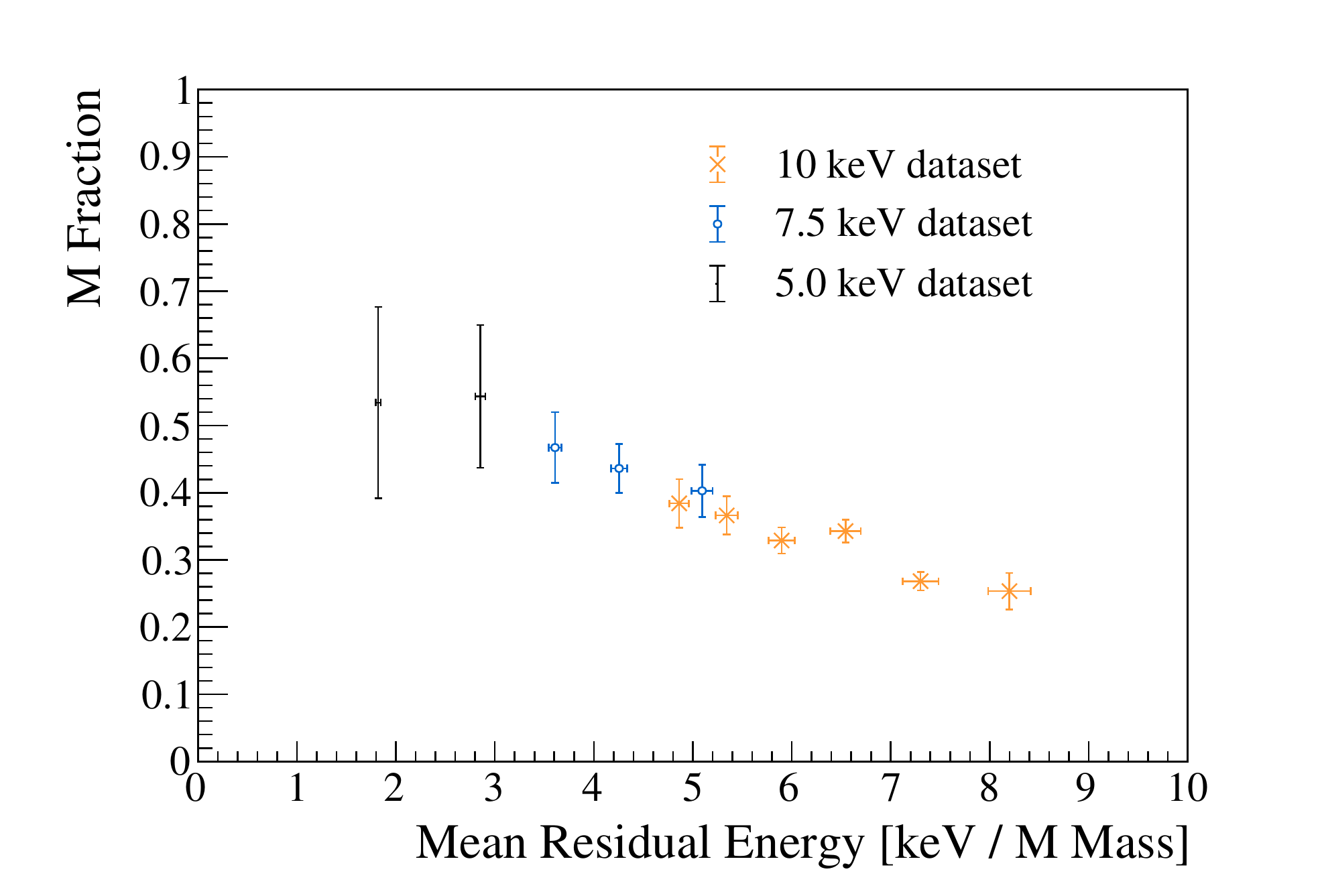}
\caption{\label{fig:mu_fraction}
M fraction measured as a function of residual energy after the foil
} 
\end{figure}

From our knowledge of the M fractions and residual energy distributions, we can determine the M conversion rate of our foil in this low incident muon energy range. The results are shown in Fig.~\ref{fig:mu_fraction}. 
The errors in the fractions are dominated by statistics, and those in the mean residual energy are correlated and arise from the uncertainty of $t_0$.
%
Our results demonstrate that in the energy range probed, a high conversion rate to M is achieved, leading to the expectation that a sizeable amount of M(2S) is also produced \cite{1981_Gabrielse}.
%

\begin{figure}[!b]
\includegraphics[width=1\columnwidth,trim={0 4 0 30},clip]{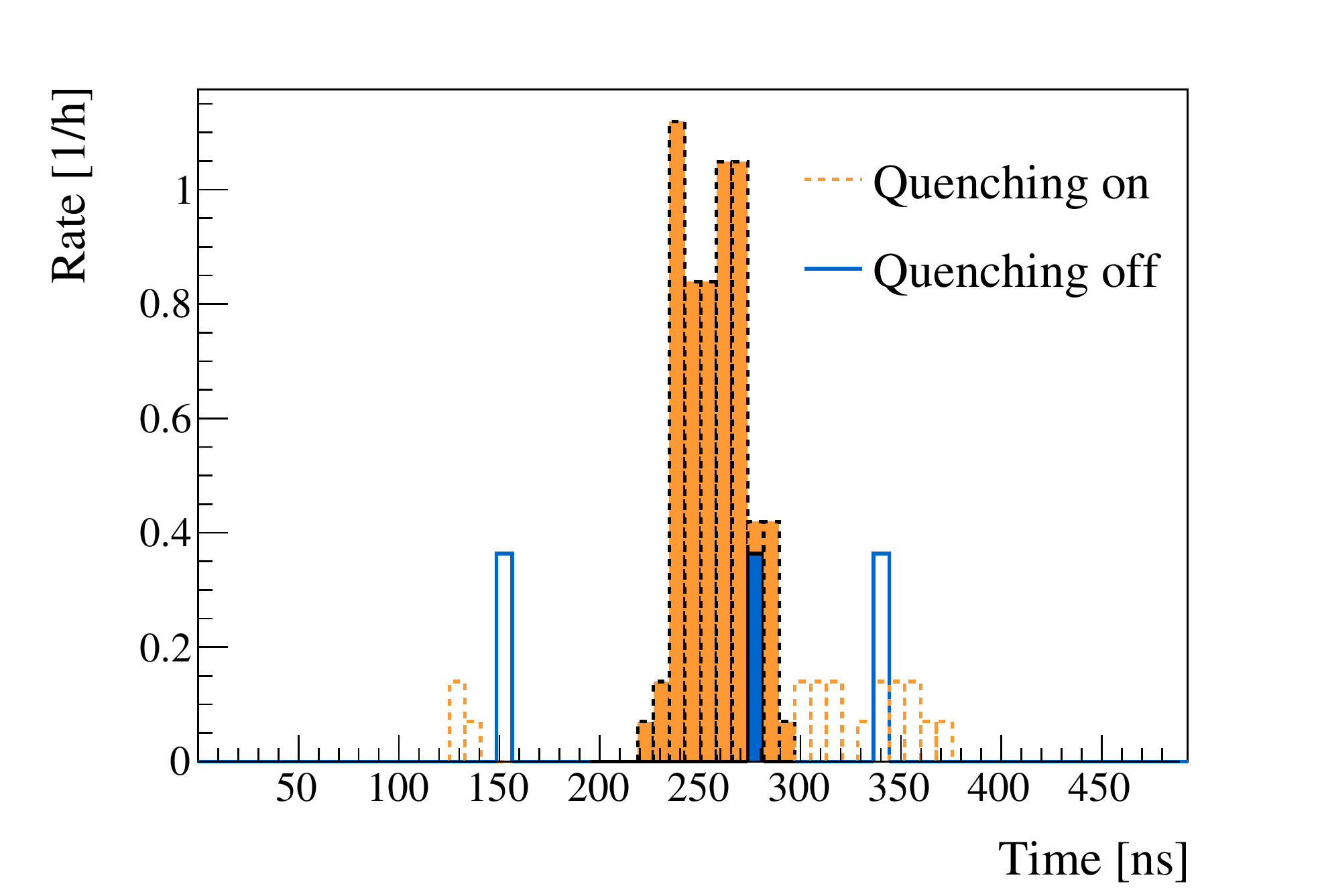}
\caption{\label{fig:mu2s}Time-of-flight distributions of the counts in the Ly-$\alpha$-MCPs, obtained from the triple coincidence dataset of $10$ keV $E_\mathrm{inc}$. The dotted data is with quenching electrodes turned on, the solid data is with quenching off. The coloured area is the time window of interest, where the Ly-$\alpha$ signal is to be expected.}
\end{figure}
%

%

%

\begin{figure}[!t]
\includegraphics[width=1\columnwidth,trim={0 4 0 0},clip]{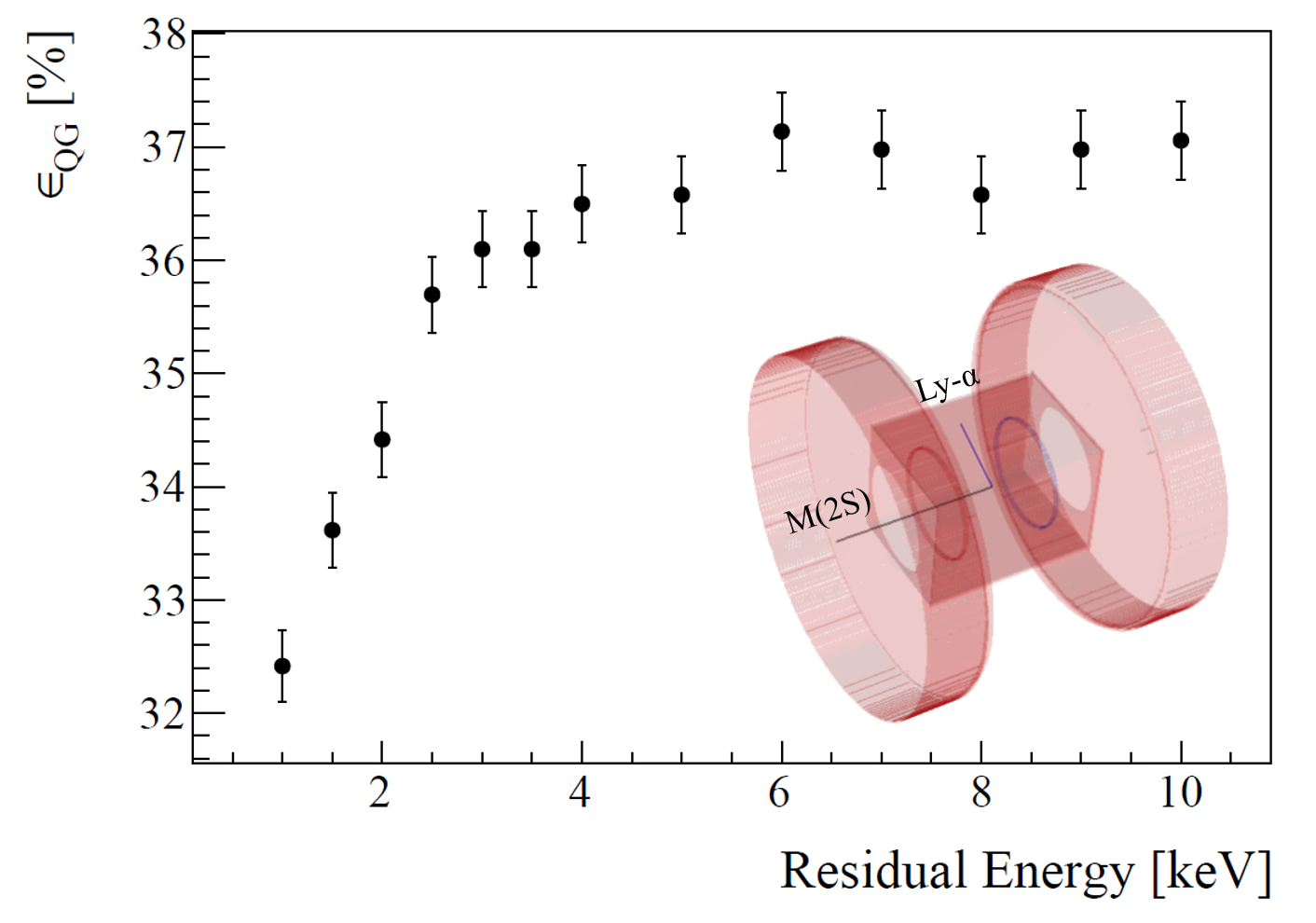}
\caption{\label{fig:quenching} Results of Monte-Carlo simulation for the quenching and geometrical efficiency as a function of energy. The inset portrays a simulated valid event where M(2S) enters the detection region, is quenched by the static field created by the two circular electrodes, and emits a photon which reached one of the detectors.}
\end{figure}

The fraction of M(2S) of the total M produced, $f_{\mathrm{2S/M}}$, is extracted from triple coincidence events between the Tag, Ly-$\alpha$, and Stop-MCPs with the quenching electrodes turned on or off, while keeping the rejection electrode turned on. The rate of triple coincidence events, $R_{\mathrm{T}}$, indicative of M(2S), is then compared to the rate of double coincidence events between the Tag and Stop-MCPs, $R_{\mathrm{D}}$, indicative of M. 
The clear triple-coincidence Ly-$\alpha$ signal is shown in Fig.~\ref{fig:mu2s} for $E_\mathrm{inc}$ of $10$ keV. The Ly-$\alpha$ signal can be seen in the expected time window calculated using the energy distributions from Fig.~\ref{fig:energy_distribution} and the distance, including the extension stage, between foil and the quenching area.
Taking into account the photon detection efficiencies, the resulting fraction of M(2S) out of the total M is
\begin{equation}
f_{\mathrm{2S/M}} = \frac{R_{\mathrm{T}}}{R_{\mathrm{D}}\cdot\epsilon_{\mathrm{QG}}\cdot\epsilon_{\mathrm{MCP}}},
\label{eq:mu_efficiency}
\end{equation}
where $\epsilon_{\mathrm{MCP}}$ stands for the Ly-$\alpha$ detection efficiency of the MCP, and $\epsilon_{\mathrm{QG}}$ for the combined efficiency for quenching as well as the solid angle covered by the detectors.
The quenching and geometrical efficiency of the Ly-$\alpha$ detection stage are correlated, and depend on the M velocity, since the position the M(2S) reaches before quenching affects the solid angle. To determine $\epsilon_{\mathrm{QG}}$, we performed a full 3D Monte-Carlo simulation of the particle motion and photon emission inside the static electric field using the SIMION 8.1 package~\cite{SIMION}. The position distribution of the particles at the detector entrance was taken from the GEANT4 beamline simulation with the calibrated foil thickness, taking into account the coincidence detection in the Stop-MCP. Additionally, the anisotropy of the photon emission relative to the electric field direction \cite{1978-Aniso}, and the transparency of the grids on the detectors, were included. The total efficiency is shown in Fig.~\ref{fig:quenching}. Folding it with the measured energy distributions, we get $\epsilon_{\mathrm{QG}}=36.4\pm0.3\%$ and $\epsilon_{\mathrm{QG}}=37.0\pm0.3\%$, for $E_\mathrm{inc}$ of $7.5$ and $10$ keV, respectively. 

The MCP detection efficiency for Ly-$\alpha$ can be estimated through $\epsilon_{\mathrm{MCP}} = \text{OAR} \cdot \epsilon_{\mathrm{CsI}}$, where OAR stands for the open-area-ratio of the MCP itself and is $0.45$ in our case. The quantum yield of the conversion from Ly-$\alpha$ to an electron in the CsI, $\epsilon_{\mathrm{CsI}}$, is in the range of $0.45-0.55$ \cite{1999-CsI,2005-Tremsin}. This leads to $\epsilon_{\mathrm{MCP}} = 0.22\pm0.02$.
The $f_{\mathrm{2S/M}}$ values, calculated according to Eq.~\ref{eq:mu_efficiency}, are summarized in Table~\ref{tab:summary} for $E_\mathrm{inc}$ of $7.5$ and $10$ keV.
Stronger scattering of the muon beam by the foil at $5$ keV $E_\mathrm{inc}$ prevented us from obtaining the reliable triple-coincidence signal needed to extract the 2S fraction.
Assuming, in accordance with hydrogen in a comparable velocity range (see Fig.~3.1 of \cite{1985-TIUMFLSThesis}), that the 2S fraction is nearly constant above $1$ keV, we obtain a weighted average value of $f_{\mathrm{2S/M}}=10\pm2 \%$.
This value agrees with estimations in the literature which span $10-13\%$ in this energy range \cite{1981-Mu2S,1984-LS,1985-TIUMFLSThesis}.

%
During the beam time, the proton current was $1.6$ mA and the corresponding rate of moderated $\mathrm{\mu^+}$ emerging was $18$ kHz. Using the LEM beamline simulation \cite{2012_MusrSim,2015-GEANT4}, with the same conditions as in our experiment, we estimate the rate of $\mathrm{\mu^+}$ passing the foil, $R_{\mathrm{\mu^+}}$, for each $E_\mathrm{inc}$.
The rate of metastable M is obtained by multiplying with the measured formation efficiencies, $R_{\mathrm{2S}} = R_{\mathrm{\mu^+}} f_{\mathrm{M/\mu^+}}  f_{\mathrm{2S/M}}$. The results are given in table \ref{tab:summary}.
We find that when increasing the beam energy, $f_{\mathrm{M/\mu^+}}$ decreases and the transmission of the beamline increases, such that the final metastable rates are comparable.
Nevertheless, the angular distribution of the beam emerging from the foil at $10$ keV is narrower, and so we concentrate on this energy for considering the rates available for a future Lamb shift experiment.

\begin{table}[!htbp]
    \sisetup{
    separate-uncertainty
    }
  \centering
  \caption{Summary of values extracted from different incident energies $E_\mathrm{inc}$. MPE is the Most Probable Energy for M that traversed the foil and reached the Stop-MCP.\\}
  \begin{ruledtabular}
    \begin{tabular}{rclccr}

    \multicolumn{1}{c}{ $E_\mathrm{inc}$} & \multicolumn{1}{c} {MPE} & \multicolumn{1}{c}{$f_{\mathrm{M/\mu^+}}$}     &  \multicolumn{1}{c}{$f_{\mathrm{2S/M}}$} & \multicolumn{1}{c} {$R_{\mathrm{\mu^+}}$} & \multicolumn{1}{c}{$R_{\mathrm{2S}}$}\\
    
     \multicolumn{1}{c} {(keV)} & \multicolumn{1}{c} {(keV)} & \multicolumn{1}{c}{(\%)}     &  \multicolumn{1}{c}{(\%)} & \multicolumn{1}{c} {(kHz)} & \multicolumn{1}{c}{(Hz)}\cr
        \noalign{\smallskip}
         \hline
        \noalign{\smallskip}
        $ 5.0~  $ & $ 2.7 \pm 0.1$  & $56.8  \pm 9.0    $ & {-}        &$1.45$ &$ 83^*   \pm 21 $ \cr
        \noalign{\smallskip}
        $ 7.5 ~ $ & $ 4.7 \pm 0.2$ & $43.2 \pm 2.4  $ & $11 \pm 4 $ &$2.07$ &$100~   \pm 30$  \cr
        \noalign{\smallskip}
        $ 10.0~ $ & $ 7.0 \pm 0.3$ & $31.8 \pm 0.8 $ & $10  \pm 3$ &$2.84$ &$ 90~ \pm   30$  \cr

%
%
    \end{tabular}%
    \end{ruledtabular}
  \label{tab:summary}%
    \begin{flushleft}* 
    For $R_{\mathrm{2S}}$ at $5$ keV, $f_{\mathrm{2S/M}}$ = $10\pm 2$ \% was assumed (see text). 
  \end{flushleft}
\end{table}%

%
%
%

In Fig.~\ref{fig:beamline}, the main missing component for precision spectroscopy experiments of the Muonium Lamb shift and fine structure is a broadband microwave apparatus that we would place in the extension stage. This could then resonantly quench the 2S beam by mixing the population with the 2P states. In this `opt-out' scheme, the Ly-$\alpha$ signal decreases near to the resonance. 

Focusing on the Lamb shift transitions, we would obtain a clean symmetric line shape of the resonance by driving the 2S $F=0 \rightarrow \mathrm{2P_{1/2}}$ $F=1$ transition around $580$ MHz. This is isolated from the next transition by $0.6$ GHz, which is favorable to that of hydrogen where the difference is only $0.2$ GHz. Based on minor improvements to the setup presented here, the expected off-resonance coincidence signal between the Ly-$\alpha$-MCP and foil is $8$/s. This rate is four orders of magnitude larger than the coincidence rate of $5$/h measured at TRIUMF \cite{1984-Epthermal}, and $4$/h at LAMPF \cite{1984-LAMPF-4h}.
%
%
To prevent the 2S $F=1$ levels from contributing to the background we would introduce a hyperfine selection stage in front of the microwave cavity in the extension section, which deexcites most of the 2S $F=1$ population to the ground state, leaving a clean beam of roughly $22$/s M(2S) $F=0$, and an off-resonance coincidence signal of $2$/s. At this rate, with $120$ hours of beamtime, the $100$ MHz natural linewidth could be resolved to $0.1$ MHz. 

%
In summary, we have demonstrated the creation of an intense directed beam of Muonium in the long-lived 2S state by transmitting low energy muons from the LEM beamline through a thin carbon foil.
With an estimation of the $\mathrm{\mu^+}$ rate as well as the measurement of the neutral and 2S fractions (see Table \ref{tab:summary}), we determined a conversion rate of $3\%$ M(2S) per incoming $\mathrm{\mu^+}$ at $10$ keV. This opens up the possibility to conduct precision measurements of laser and microwave transitions from the M(2S) state. For a measurement of the $n=2$ Lamb shift, arguably the most promising of these transitions, an uncertainty on the order of $100$ kHz is projected, which constitutes an improvement by two orders of magnitude over the best determination from the literature \cite{1984-LS}. 

A determination of the Lamb shift in muonium with this accuracy will provide a stringent test of high-order recoil corrections in bound state QED \cite{1998-JungLamb,2019-Pahucki}, free of finite-size effects \cite{2005-PSAS}. Moreover, it will be a sensitive probe for the existence of exotic dark-sector particles \cite{2019-Peset}, new muonic forces \cite{2011-MuForce}, and hidden dimensions \cite{2008-Dim,2016-Dim}.

The work presented here is supported by the European Research Council (grant number 818053-Mu-MASS), the Yamada Science Foundation, and ETH Z\"urich (grant number ETH-46 17-1). BO is supported by the Israel Academy of Sciences and Humanities. We would like to acknowledge the contributions of D. Nandal, A. Nanda and J. Zhang during various stages of this work. The muon measurements have been performed at the Swiss Muon Source S$\mu$S, Paul Scherrer Institute, CH-5232 Villigen, Switzerland.


%
\bibliographystyle{apsrev4-1}
\bibliography{sample.bib}

\end{document}